\documentclass[10pt,conference,letterpaper]{IEEEtran}

\usepackage{cite}
\usepackage{psfrag}
\usepackage{multirow}

\ifCLASSINFOpdf
  \usepackage[pdftex]{graphicx}
  \graphicspath{{./pdf/}{./jpeg/}{./figs/}}
  \DeclareGraphicsExtensions{.pdf,.jpeg,.png}
\else
  \usepackage[dvips]{graphicx}
  \graphicspath{{./figs/}}
  \DeclareGraphicsExtensions{.eps}
\fi

\usepackage[cmex10]{amsmath}
\usepackage{amssymb} 
\usepackage{exscale,relsize}

\DeclareMathSizes{9}{8}{7}{6}
\DeclareMathSizes{8}{7}{6}{5}

\usepackage{url}

\usepackage{verbatim}

\newtheorem{definition}{Definition}
\newtheorem{proposition}{Proposition}

\IEEEoverridecommandlockouts

\begin{document}
\title{QBF-Based Boolean Function Bi-Decomposition}

\author{Huan Chen\\
University College Dublin\\
Dublin, Ireland\\
huan.chen@ucd.ie\\
\hspace{-2.5cm}
\and
Mikol\'{a}\v{s} Janota\\
INESC-ID\\
Lisbon, Portugal\\
mikolas.janota@gmail.com\\
\hspace{-2.5cm}
\and
Joao Marques-Silva\\
Unversity College Dublin\\
Dublin, Ireland\\
jpms@ucd.ie\\
}

\maketitle

\label{abs}
\begin{abstract}
This paper is an extension of~\cite{ChenDATE12}.
Boolean function bi-decomposition is ubiquitous in logic synthesis. It 
entails the decomposition of a Boolean function using two-input simple
logic gates.
Existing solutions for bi-decomposition are often based on BDDs and,
more recently, on Boolean Satisfiability.
In addition, the partition of the input set of variables is either
assumed, or heuristic solutions are considered for finding good
partitions.
In contrast to earlier work, this paper proposes the use of Quantified
Boolean Formulas (QBF) for computing bi-decompositions. These
bi-decompositions are optimal in terms of the achieved disjointness
and balancedness of the input set of variables.
Experimental results, obtained on representative benchmarks,
demonstrate clear improvements in the quality of computed
decompositions, but also the practical feasibility of QBF-based
bi-decomposition.
%
%
%
\end{abstract}
%




\section{Introduction}
\label{01-introduction}
%
Boolean function decomposition is a fundamental technique in 
logic synthesis.
Given a complex Boolean function $f(X)$, function decomposition
consists of representing $f(X)$ as $f(X) = h(g_1(X), \dots, g_m(X))$,
often with $m < ||X||$, such that $h, g_1, \dots, g_m$ are simpler
sub-functions. 
%
%
Boolean function decomposition plays an important role in modern
Electronic Design Automation (EDA), including multi-level logic
synthesis and FPGA
synthesis~\cite{LaiDAC93,LubaVLSID95,SchollFPGA01}.

Bi-decomposition~\cite{MalikICCD91,BochmannEDAC91,StanionARVLSI95,SasaoIWLS97,MishchenkoDAC01,CortadellaTCAD03,LeeDAC08,ChoudhuryICCAD10,ChenVLSISOC11},
a special form (with $m=2$) of functional decomposition, is arguably the
most widely used form of Boolean function decomposition.
Bi-decomposition consists of  decomposing Boolean function $f(X)$ into
the form of $f(X) = h(f_A(X_A,X_C),f_B(X_B,X_C))$, under variable
partition $X = \{X_A|X_B|X_C\}$ wherein fewer number of variables are
required in each sub-set $X_A$, $X_B$ and $X_C$.
%
%

The quality of Boolean function decomposition is often related with
the quality of variable
partitions~\cite{LeeDAC08,LinICCAD08,ChoudhuryICCAD10,ChenVLSISOC11}, 
as an optimal solution requires fewer input variables and simpler
sub-functions.
%
%

Similar to recent work~\cite{LeeDAC08,LinICCAD08,ChenVLSISOC11}, this
paper addresses two {\em relative} metrics measuring the quality of
bi-decompositions, namely {\em disjointness} and {\em balancedness}.
%
In practice, disjointness is in general preferred~\cite{LeeDAC08}, 
since it reduces the number of shared input variables between $f_A$
and $f_B$. In turn, this often reduces complexity of the resulting
Boolean network.
{\em Absolute} quality metrics are an alternative to relative quality
metrics, and include total variable count ($\Sigma$) and maximum 
partition size ($\Delta$)~\cite{ChoudhuryICCAD10}. Nevertheless,
absolute quality metrics scale worse with the number of inputs~\cite{ChoudhuryICCAD10}.

Decomposition of Boolean functions has been extensively studied, and
initial work can be traced back to
1950s~\cite{AshenhurstISTS57,Curtis62}.
The very first algorithm for bi-decomposition was presented for the AND case in~\cite{MalikICCD91}.
The first solution for XOR case was given in~\cite{SasaoIWLS95}.
The general case of bi-decomposing of boolean network was proposed in the work~\cite{StanionARVLSI95}.
Traditional
approaches~\cite{LaiDAC93,ChangTCAD96,MishchenkoDAC01,SchollFPGA01,CortadellaTCAD03}
use BDDs as the underlying data structure.
However, BDDs impose severe constraints on the number of input
variables circuits can have. It is also generally accepted that BDDs
do not scale for large Boolean functions.
As a result, recent
work~\cite{LeeDAC08,LinICCAD08,JiangTComp10,ChenVLSISOC11} proposed
the use of Boolean Satisfiability (SAT) and Minimally Unsatisfiable
Subformulas (MUS) to manipulate large Boolean functions.
This resulted in significant performance improvements. In addition,~\cite{LeeDAC08,ChenVLSISOC11} proposed heuristic approaches for identifying variable
partitions.
Explicit (but heuristically restricted) enumeration of variable
partitions~\cite{LeeDAC08,LinICCAD08,JiangTComp10} sometimes produces
good solutions, corresponding to adequate values of disjointness and
balancedness.
However, it is in general difficult to guarantee the quality of
variable partitions, since the number of possible partitions grows
exponentially with the number of inputs. This prevents brute-force
search~\cite{LeeDAC08} in practice.
%

%

This paper addresses the problem of computing bi-decompositions with
optimum variable partitions. The optimality of achieved variable
partitions is measured in terms of existing metrics, namely
disjointness and balancedness. The proposed solutions are based on
novel QBF formulations for the problem of Boolean function
bi-decomposition subject to target metrics (e.g. disjointness,
balancedness, etc.). Besides the novel QBF formulations, the paper
shows how bi-decomposition can be computed with {\em optimum} values
for the target metrics.
Experimental results, obtained on well-known benchmarks, demonstrate
that QBF-based function bi-decomposition performs comparably with
recent heuristic
approaches~\cite{LeeDAC08,LinICCAD08,JiangTComp10,ChenVLSISOC11},
while guaranteeing optimum variable partitions.

%
%
%

The paper is organized as follows.
Section~\ref{02-preliminaries} covers the preliminaries.
Section~\ref{03-related-work} reviews models for Boolean function bi-decomposition.
Section~\ref{04-new-model} proposes the new QBF-based models.
Section~\ref{06-experimental-results} presents the experimental results.
Finally, section~\ref{07-conclusion} concludes the paper and outlines future work.
%


\section{Preliminaries}
\label{02-preliminaries}
%
Variables are represented by set $X = \{x_1,x_2,\dots,x_n\}$.
The cardinality of $X$ is denoted as $||X||$.
A partition of a set $X$ into $X_i \subseteq X$ for $i = 1,\dots,k$ (with $X_i \bigcap X_j = \emptyset, i \neq j$ and $\bigcup_iX_i = X$) is denoted by $\{X_1|X_2|\dots|X_k\}$.
A Completely Specified Function (CSF) is denoted by $f : \mathcal{B}^n \rightarrow \mathcal{B}$.
Similar to the recent work~\cite{LeeDAC08,ChenVLSISOC11}, this paper assumes CSFs.
%
%
\subsection{Boolean Function Bi-Decomposition}
\begin{definition}
{\em
Bi-decomposition~\cite{SasaoIWLS97} for Completely Specified Function
(CSF) $f(X)$ consists of decomposing $f(X)$ under variable partition $X =
\{X_A|X_B|X_C\}$, into the form of $f(X) = f_A(X_A,X_C)$ $<$OP$>$
$f_B(X_B,X_C)$, where $<$OP$>$ is a binary operator, typically {\em
  OR}, {\em AND} or {\em XOR}. 
}
\end{definition}

This paper addresses {\em OR}, {\em AND} and {\em XOR}
bi-decomposition because these three basic gates form other types of
bi-decomposition~\cite{LeeDAC08}.
%
%
Bi-decomposition is termed {\em disjoint} if $||X_C|| = 0$.
A partition of $X$ is trivial if $X = X_A \bigcup X_C$ or $X = X_B \bigcup X_C$ holds.
Similar to earlier work~\cite{LeeDAC08,LinICCAD08,ChenVLSISOC11}, this paper addresses non-trivial bi-decompositions.
%
%
\subsection{Boolean Satisfiability}
Boolean formulas $\phi$ and $\psi$ are defined over a finite set of Boolean variables $X$.
Individual variables are represented by lowercase letters $x$, $y$, $z$, $w$ and $o$, and subscripts may be used (e.g. $x_1$).
The Boolean connectives considered will be $\neg$, $\rightarrow$, $\leftrightarrow$, $\wedge$, $\vee$. When necessary, parentheses are used to enforce precedence. 
A formula in Conjunctive Normal Form (CNF) $\mathcal{F}$ is defined as a set of sets of literals defined on $X$,
representing a conjunction of disjunctions of literals.
A literal is either a variable or its complement.
Each set of literals is referred to as a clause $c$.
Moreover, it is assumed that each clause is non-tautological.
Additional SAT definitions can be found in standard references (e.g.~\cite{SATBook09}).
\subsection{Quantified Boolean Formulas}
Quantified Boolean Formulas (QBF) generalize Boolean formulas by
quantifying variables, either existentially ($\exists$) or universally
($\forall$). 
Variables in a QBF being quantified are referred to as {\em bound}
variables, whereas those not quantified are referred to as {\em free}
variables.
Throughout this paper, all Boolean variables in QBF are assumed to be
quantified (bound).
QBF are assumed to be in the {\em prenex} form $Q_1p_1 \dots
Q_np_n. \phi$, with $Q_i \in \{ \exists, \forall \}$,
$p_i$ are distinct Boolean variables, and $\phi$ is a propositional
formula using only the variables $p_i$ and the constants $0$ (false),
$1$ (true).
For example, in a QBF $\exists x, \forall z. f(x,y,z)$, $x$ and $z$
are bound variables and $y$ is a free variable, prefix is $\exists x,
\forall z$ and matrix is $f(x,y,z)$.
%
%

The problem of solving a QBF is referred to as quantified satisfiability (QSAT).
The general case of a QSAT problem is PSPACE-complete~\cite{SATBook09},
which is computationally hard compared to Boolean SAT problems.
Restrictions in the number of quantifier alternations characterize 
the polynomial hierarchy~\cite{SATBook09}.
QBF formulas with $k$ quantifier alternations, that start with
$\exists$ are denoted QBF$_{k,\exists}$, whereas those that start with
$\forall$ are denoted QBF$_{k,\forall}$. Formulas in QBF$_{k,\exists}$
are $\Sigma_k^P$-complete, where formulas in QBF$_{k,\forall}$ are
$\Pi_k^P$-complete.
%
%
%
%
\subsection{Quality Metrics}
The quality of variable partitions mainly impacts the quality of
bi-decomposition~\cite{LeeDAC08,ChoudhuryICCAD10,ChenVLSISOC11}, and
indirectly impacts the decomposed network, e.g. delay, area and
power consumption~\cite{ChoudhuryICCAD10}. 
Similar to~\cite{LeeDAC08,LinICCAD08,ChenVLSISOC11}, this paper
measures the quality of variable partitions through two {\em relative}
quality metrics, namely {\em disjointness} and {\em balancedness}.
Assume a variable partition $\{X_A|X_B|X_C\}$ for $f(X)$, where $X_A$, $X_B$ and $X_C$ are the sets of the input variables to decomposition functions $f_A$, $f_B$ and common to $f_A$ and $f_B$, respectively.
\begin{definition}
[Disjointness]{$\epsilon_D = \frac{||X_C||}{||X||}$ denotes the ratio of the number of common variables to inputs.
A value of $\epsilon_D$ close to 0 is preferred, as $\epsilon_D = 0$
represents a disjoint bi-decomposition.}
\end{definition}
\begin{definition}
[Balancedness]{$\epsilon_B = \frac{\big|||X_A||-||X_B||\big|}{||X||}$ denotes the absolute size difference between $X_A$ and $X_B$.
$\epsilon_B = 0$ represents a balanced variable partition.}
\end{definition}

In practice, disjointness is preferred since a lower value represents
a smaller number of shared input variables of the resulting decomposed
circuit that typically has smaller area and power footprint.
A lower balancedness typically corresponds to smaller delay of the
decomposed network.

This paper develops QBF models for achieving disjointness and
balancedness targets, or even the optimum solutions.

%
%


\section{Related Work}
\label{03-related-work}
%
Bi-Decompositions of Boolean functions are either based on Binary
Decision Diagrams (BDDs) or on Boolean Satisfiability (SAT). 
This section briefly overviews earlier work.

\subsection{BDD-Based Bi-Decomposition}
BDDs are a canonical representation of Boolean functions, and have
been widely applied in function
decomposition~\cite{LaiDAC93,ChangTCAD96,MishchenkoDAC01,SchollFPGA01,CortadellaTCAD03}. Variants
of BDDs have also been considered~\cite{ChoudhuryICCAD10}, targeting
optimization of area~\cite{MishchenkoDAC01,ChoudhuryICCAD10},
delay~\cite{ChangTCAD96,MishchenkoDAC01,CortadellaTCAD03,ChengTCAD08,ChoudhuryICCAD10},
and power consumption~\cite{ChoudhuryICCAD10}. 
Algorithms based on BDDs have several advantages, including flexible
Boolean function manipulation~\cite{SasaoIWLS97,MishchenkoDAC01},
on-demand selection of variable
partitions~\cite{KravetsDATE09,ChoudhuryICCAD10}, and the ability to
handle don't-care conditions~\cite{MishchenkoDAC01,ChengTCAD08}.
%
%
Algorithms based on BDDs also have key drawbacks. BDD-based approaches
are generally memory intensive, sensitive to variable
orders~\cite{ChoudhuryICCAD10}, restricted on the number of primary
inputs~\cite{LeeDAC08}, and on the number of variable partitions
considered. 

\subsection{SAT-Based and MUS-Based Bi-Decomposition}
With the objective of target bi-decomposition of large Boolean
functions with a large number of inputs, and correspondingly large
number of variable partitions, recent work proposed SAT-based
bi-decomposition~\cite{LeeDAC08} and MUS-based
bi-decomposition~\cite{LeeDAC08,ChenVLSISOC11}.
SAT-based OR, AND and XOR bi-decompositions under known and unknown
partition of variables were proposed in~\cite{LeeDAC08}.
For example, the widely used OR bi-decomposition can be computed by
SAT solving~\cite{LeeDAC08}.
Given a non-trivial variable partition $X=\{X_A|X_B|X_C\}$, the
following result holds:
\begin{proposition}
{\em
~\cite{LeeDAC08} A completely specified function $f(X)$ can be written as $f_A(X_A,X_C) \vee f_B(X_B,X_C)$ for some functions $f_A$ and $f_B$ if and only if the Boolean formula
%
%

\begin{equation}
\label{E_sat}
\textstyle
f(X_A, X_B, X_C) \wedge \neg f(X_A',X_B,X_C) \wedge \neg f(X_A,X_B',X_C)
\end{equation}
\normalsize
is unsatisfiable, where variable set $Y'$ is an instantiated version of variable set $Y$.
}
\end{proposition}
An {\em instantiated} version $x'$ of Boolean variable $x$ can be viewed as a new Boolean variable $x'$ that replaces $x$.
This approach assumes that a variable partition $X = \{X_A|X_B|X_C\}$ is given.
In practice, such variable partitions are generally unknown and must be automatically derived.
One possible approach is to consider the following
formulation~\cite{LeeDAC08}:
%
%

\begin{equation}
\label{E_sat_unknown}
\begin{split}
f(X) \wedge \neg f(X') \wedge \bigwedge_{i} ((x_i \equiv x'_i) \vee \alpha_{x_i})\\
\wedge \neg f(X'') \wedge \bigwedge_{i} ((x_i \equiv x''_i) \vee \beta_{x_i})
\end{split}
\end{equation}
\normalsize
where $x' \in X'$ and $x'' \in X''$ are instantiated versions of $x \in X$.
$\alpha_{x_i}$ and $\beta_{x_i}$ are {\em control} variables for enumerating variable partitions.
By assigning different Boolean values to $\alpha_{x_i}$ and $\beta_{x_i}$, some of the clauses $((x_i \equiv x'_i) \vee \alpha_{x_i})$, $((x_i \equiv x''_i) \vee \beta_{x_i})$ are relaxed.
The resulting clauses $(x_i \equiv x_i')$ and $(x_i \equiv x_i'')$
impose equivalence relations for each pair of variables in sets $X$
and $X'$, and in $X$ and $X''$, respectively.

The original work on SAT-based bi-decomposition~\cite{LeeDAC08} proposed the use of interpolation for computing the target functions $f_A$ and $f_B$.
Given that our work focuses on improving the identification of variable partitions, interpolation can also be used for computing functions $f_A$ and $f_B$.
Similarly to OR bi-decomposition, AND and XOR bi-decomposition can be
computed by using SAT.
Due to space limitations, this section omits the explanation of SAT-based AND and XOR bi-decompositions (e.g.~see \cite{LeeDAC08}).
The approaches proposed in~\cite{LeeDAC08} are referred to as {\bf
  {\em LJH}} in the remainder of the paper.

SAT-based bi-decomposition~\cite{LeeDAC08,ChenVLSISOC11} proposed a
number of MUS-based techniques for computing good variables
partitions. These include plain MUS computation and, more recently,
group-oriented MUS computation. These approaches can be viewed as 
practical engineering solutions for bi-decomposition of {\em large}
Boolean functions.
Nevertheless, these approaches are heuristic and provide no guarantees
regarding the quality of computed variable partitions.

%
%


\section{QBF-Based Bi-Decomposition}
\label{04-new-model}
%
This section develops QBF models for computing Boolean function
bi-decomposition with optimum variable partitions.
The case for OR bi-decomposition is considered first. Afterward, the
paper summarizes AND and XOR bi-decomposition.

\subsection{OR Bi-Decomposition}

Observe that, as described above, assignments to the $\alpha_{x_i}$
and $\beta_{x_i}$ variables specify the sets $X_A$, $X_B$ and $X_C$.
A key observation is that formulation (\ref{E_sat_unknown}) above
provides the basis for a natural (albeit incomplete) QBF formulation.
Formulation (\ref{E_sat_unknown}) essentially quantifies existentially
the $\alpha_{x_i}$ and $\beta_{x_i}$ variables and, by requiring
unsatisfiability, quantifies the $X, X', X''$ variables universally. 
%
%
This results in the following QBF formulation:
%
%

\begin{equation}
\label{E_2qbf_basic}
\begin{split}
\exists_{\alpha_{x_i},\beta_{x_i}},\forall_{X,X',X''}.
\neg [f(X) \wedge \neg f(X') \wedge \bigwedge_{i} ((x_i \equiv x'_i) \vee \alpha_{x_i})\\
\wedge \neg f(X'') \wedge \bigwedge_{i} ((x_i \equiv x''_i) \vee \beta_{x_i})]
\end{split}
\end{equation}
\normalsize
This QBF formulation has a few important drawbacks: (i) the solution
can be a trivial partition; and (ii) the quality of a non-trivial
partition can be arbitrary.
As a result, the ability to control the quality of the computed
variable partition, requires extending (\ref{E_2qbf_basic}) as
follows:
%
%

\begin{equation}
\label{E_2qbf_unknown}
\begin{split}
\exists_{\alpha_{x_i},\beta_{x_i}},\forall_{X,X',X''}.
\neg [f(X) \wedge \neg f(X') \wedge \bigwedge_{i} ((x_i \equiv x'_i) \vee \alpha_{x_i})\\
\wedge \neg f(X'') \wedge \bigwedge_{i} ((x_i \equiv x''_i) \vee \beta_{x_i})]\\
\wedge f_N(\alpha_X,\beta_X) \wedge f_T(\alpha_X,\beta_X)
\end{split}
\end{equation}
\normalsize
where $f_N(\alpha_X,\beta_X)$ requires a non-trivial variable
partition, and $f_T(\alpha_X,\beta_X)$ requires the computed
variable partition to respect target metrics, e.g.~disjointness or
balancedness.

\subsubsection{Ensuring Non-trivial Partitions}
Filtering of trivial partitions is achieved through constraints added
to $f_N(\alpha_X,\beta_X)$.
A {\em trivial partition} of $X$ is such that either $X = X_A \bigcup
X_C$ or $X = X_B \bigcup X_C$ holds. In other words, a {\em
  non-trivial} partition is such that $X_A \neq \emptyset \wedge X_B
\neq \emptyset$.
%
This condition is expressed with cardinality constraints
$AtLeast1(\bigcup_{x \in X} \alpha_x) \wedge AtLeast1(\bigcup_{x \in
  X} \beta_x)$.

\subsubsection{Targeting Disjointness}

Constraints on variables $\alpha_{x_i}$ and $\beta_{x_i}$ serve to
require target values of disjointness.
Observe that in model~\eqref{E_2qbf_unknown}, the assignment
$(\alpha_x,\beta_x) = (0,0)$ denotes that $x \in X_C$. Improvements in
disjointness consists of reducing the size of $X_C$. This is achieved
by computing variable partitions with a sufficiently small number of
pairs $(\alpha_x,\beta_x)$ with $(\alpha_x,\beta_x) = (0,0)$.
For a target level of disjointness $\epsilon$, with $0\le \epsilon <
1$, let $k \in \mathbb{N}, k = \lfloor ||X|| \cdot \epsilon \rfloor$.
Hence, the target constraint $f_T(\alpha_X,\beta_X)$ is defined as
follows:
%
%

\begin{equation}
\label{E_card_dis}
(\sum_{x \in X} \overline{\alpha_x} \cdot \overline{\beta_x}) \leq k
\end{equation}
\normalsize
Clearly, $k$ is discrete and finite, and so the optimum value can be
computed by iteratively solving QBF~(\ref{E_2qbf_unknown}) for
different values of $k$.

Moreover, observe that Boolean function bi-decomposition exhibits key
symmetry properties. For example, sets $X_A$ and $X_B$ are
{\em indistinguishable}, and so the optimum solution is obtained even
if constraint $||X_A||\ge ||X_B||$ is included in the problem 
formulation. This constraint can either be added to
$f_N(\alpha_X,\beta_X)$ or to $f_T(\alpha_X,\beta_X)$.
In practice, 
this optimization reduces substantially the search space of the
resulting QBF.

\subsubsection{Targeting Balancedness}
Similarly to the approach for disjointness, constraints on variables
$\alpha_{x_i}$ and $\beta_{x_i}$ serve to require target values of
balancedness.
Balancedness is improved if the difference between the number of
variables $x$ in set $X_A$, i.e. $(\alpha_x,\beta_x) = (1,0)$, and the
number of variables $x$ in set $X_B$, i.e. $(\alpha_x,\beta_x) =
(0,1)$ is {\em minimized}.
For a target level of balancedness $\epsilon$, with $0\le \epsilon <
1$, let $k \in \mathbb{N}, k = \lfloor ||X|| \cdot \epsilon \rfloor$.
Hence, the target constraint $f_T(\alpha_X,\beta_X)$ is defined as
follows:
%
%

\begin{equation}
\label{E_card_bal}
0 \leq (\sum_{x \in X} \alpha_x \cdot \overline{\beta_x} - \sum_{x \in
  X} \overline{\alpha_x} \cdot \beta_x) \leq k
\end{equation}
\normalsize
Observe that, as before, the optimum value of $k$ can be searched for,
by iteratively solving the QBF~(\ref{E_2qbf_unknown}) for different
values of $k$.
In addition, note that, in this case, the symmetry between $X_A$ and
$X_B$ is automatically removed, by requiring $||X_A||\ge ||X_B||$.

\subsubsection{Integrating Disjointness \& Balancedness}
In practical settings, it is often the case that the objective is to
achieve some simultaneous level of disjointness and balancedness.
\begin{definition}[Cost of Disjointness and Balancedness]
The cost of Disjointness and Balancedness is the arithmetic sum of
weighted Disjointness and weighted Balancedness, which is expressed as
the following cost function:
%

\begin{equation}
\label{E_cost_DB}
\sum \varpi_D \cdot Disjointness + \varpi_B \cdot Balancedness
\end{equation}
\normalsize
where $\varpi_D$ ($\varpi_D \in [0,1]$) and $\varpi_B$ ($\varpi_B \in
[0,1]$) are weights for Disjointness and Balancedness, respectively.
\end{definition}

Observe that cost function (\ref{E_cost_DB}) can be simplified if
disjointness and balancedness are equally preferred, i.e.~$\varpi_D =
\varpi_B = 1$.
Moreover, if $||X_A||$ is assumed to be no less than $||X_B||$, then
the cardinality constraint can be simplified as follows.
%
%
%

\begin{equation}
\label{E_card_sum}
0 \leq (\sum_{x \in X} \overline\alpha_x \cdot \overline{\beta_x} + \sum_{x \in X} \alpha_x \cdot \overline{\beta_x} - \sum_{x \in X} \overline{\alpha_x} \cdot \beta_x) \leq k
\end{equation}
\normalsize
where $k \in \mathbb{N}, k = \lfloor ||X|| \cdot \epsilon \rfloor$.
%
%
\subsubsection{Practical Implementation}
In practice, the use of the 2QBF formula~\eqref{E_2qbf_unknown} is not
straightforward because it requires auxiliary variables to encode
it into CNF, as required by most QBF solvers.
These auxiliary variables are existentially quantified in the
innermost level of the QBF prefix and consequently result in a 3QCNF 
formula.

Consider a 2QBF formula $\exists_Z,\forall_X.\phi$, where $\phi$ is
not in CNF. As indicated above, converting $\phi$ to CNF requires
additional variables, which results in a 3QCNF formula.
Instead of using a solver for QBF formulas with three levels of
quantifiers, a different approach is used, which has been recently 
suggested in~\cite{JanotaSAT11}.
Consider the negation of $\exists_Z,\forall_X.\phi$,
i.e.~$\forall_Z,\exists_X.\neg\phi$. 
Observe that if $\exists_Z,\forall_X.\phi$ is valid, then 
$\forall_Z,\exists_X.\neg\phi$ cannot be valid. Thus, if the
QBF solver provides a {\em counterexample} for why
$\forall_Z,\exists_X.\neg\phi$ cannot be satisfied, it represents a
model of $\exists_Z,\forall_X.\phi$. For QBF~(\ref{E_2qbf_unknown}),
the model represents the intended variable partition.
As a result, the 2QBF formula to be used becomes:
%

\begin{equation}
\begin{split}
\label{E_2qbf_unknown_new}
\forall_{\alpha_{x_i},\beta_{x_i}},\exists_{X,X',X''}.
[f(X) \wedge \neg f(X') \wedge \bigwedge_{i} ((x_i \equiv x'_i) \vee \alpha_{x_i})\\
\wedge \neg f(X'') \wedge \bigwedge_{i} ((x_i \equiv x''_i) \vee \beta_{x_i})]\\
\vee \neg f_N(\alpha_X,\beta_X) \vee \neg f_T(\alpha_X,\beta_X)
\end{split}
\end{equation}
\normalsize

\subsubsection{Finding the Optimum}
This section summarizes the approaches that can be used for computing
the optimum disjointness or balancedness.
An initial upper bound, on both disjointness and balancedness, can be
obtained with the group-oriented MUS-based
model~\cite{ChenVLSISOC11}. Alternatively, the upper bound can be set
to 1.
Three strategies have been studied for computing the optimum
disjointness and balancedness values.

Monotonically Increasing (MI) denotes iteratively increasing the value
of $k$. Monotonically Decreasing (MD) denotes iteratively decreasing
the value of $k$. Finally, dichotomic divide-and-conquer denotes
binary search (Bin).
In our experiments, the best results for disjointness were obtained
using the sequence: MD $\rightarrow$ Bin $\rightarrow$ MI, where the
number of iterations for each is heuristically chosen. For
balancedness the best results for balancedness were obtained with MI.

\subsection{AND/XOR Bi-Decomposition}
AND bi-decomposition is the dual of OR bi-decomposition and can be
converted from the construction of OR
bi-decomposition~\cite{MishchenkoDAC01,LeeDAC08,ChenVLSISOC11}. 
The proposed QBF model~\eqref{E_2qbf_unknown_new} is able to decompose
$\neg f$ into $f_A \vee f_B$.
By negating both sides, $f$ is decomposed into $\neg f_A \wedge \neg
f_B$~\cite{LeeDAC08}.
QBF-based XOR bi-decomposition is similar to OR
bi-decomposition~\cite{LeeDAC08,ChenVLSISOC11} and can be explained
with an analogous derivation of the model~\eqref{E_2qbf_unknown_new}.
The full derivation of QBF-based AND/XOR bi-decomposition is omitted
due to lack of space.
\begin{table*}[!t]
\caption{Comparison of quality metrics between OR models}
\centering
\scalebox{0.7}{
\begin{tabular}{|l|r|r|r|r|r|r|r|r|r|r|r|r|r|r|r|}
\hline
\hline
\multicolumn{1}{|c|}{\multirow{4}{*}{Circuit}} & \multicolumn{3}{c|}{\multirow{2}{*}{Circuit Statistics}} & \multicolumn{6}{c|}{$\mathcal{OR}$ LJH~\cite{LeeDAC08} vs. STEP-\{QD,QB,QDB\}} & \multicolumn{6}{c|}{$\mathcal{OR}$ STEP-MG~\cite{ChenVLSISOC11} vs. STEP-\{QD,QB,QDB\}}\\
\cline{5-16}
& \multicolumn{3}{c|}{} & \multicolumn{2}{c|}{Disjointness} & \multicolumn{2}{c|}{Balancedness} &  \multicolumn{2}{c|}{Disjointss+Balancedness} & \multicolumn{2}{c|}{Disjointness} & \multicolumn{2}{c|}{Balancedness} &  \multicolumn{2}{c|}{Disjointness+Balancedness}\\
\cline{2-16}
& \multirow{2}{*}{\#In{~}} & \multirow{2}{*}{\#InM} & \multirow{2}{*}{\#Out} & STEP-QD & Both two are & STEP-QB & Both two are & STEP-QDB & Both two are &  STEP-QD & Both two are & STEP-QB & Both two are & STEP-QDB & Both two are\\
& & & & better (\%) & equal (\%) & better (\%) & equal (\%) & better (\%) & equal (\%) & better (\%) & equal (\%) & better (\%) & equal (\%) & better (\%) & equal (\%)\\
\hline
\hline
C7552 & 207 & 194 & 108 & 30.00 & 70.00 & 50.00 & 50.00 & 0.00 & 100.00 & 0.00 & 100.00 & 35.29 & 64.71 & 16.67 & 83.33\\
s15850.1 & 611 & 183 & 684 & 0.00 & 100.00 & 7.69 & 92.31 & 7.69 & 92.31 & 16.61 & 83.39 & 52.10 & 47.90 & 14.09 & 85.91\\
s38584.1 & 1464 & 147 & 1730 & 18.60 & 81.40 & 70.15 & 29.85 & 41.76 & 58.24 & 23.14 & 76.86 & 84.54 & 15.46 & 41.12 & 58.88\\
C2670 & 233 & 119 & 140 & 8.33 & 91.67 & 48.72 & 51.28 & 11.54 & 88.46 & 22.22 & 77.78 & 76.92 & 23.08 & 42.31 & 57.69\\
i10 & 257 & 108 & 224 & 17.57 & 82.43 & 73.23 & 26.77 & 18.92 & 81.08 & 50.51 & 49.49 & 84.87 & 15.13 & 21.62 & 78.38\\
s38417 & 1664 & 99 & 1742 & 12.28 & 87.72 & 52.65 & 47.35 & 6.45 & 93.55 & 20.62 & 79.38 & 60.94 & 39.06 & 4.94 & 95.06\\
s9234.1 & 247 & 83 & 250 & 11.58 & 88.42 & 60.78 & 39.22 & 8.11 & 91.89 & 14.15 & 85.85 & 71.30 & 28.70 & 18.60 & 81.40\\
rot & 135 & 63 & 107 & 4.17 & 95.83 & 72.92 & 27.08 & 8.33 & 91.67 & 33.87 & 66.13 & 87.10 & 12.90 & 30.56 & 69.44\\
s5378 & 199 & 60 & 213 & 10.38 & 89.62 & 82.24 & 17.76 & 5.00 & 95.00 & 17.27 & 82.73 & 90.09 & 9.91 & 15.00 & 85.00\\
s1423 & 91 & 59 & 79 & 3.85 & 96.15 & 42.31 & 57.69 & 5.00 & 95.00 & 21.21 & 78.79 & 51.61 & 48.39 & 0.00 & 100.00\\
pair & 173 & 53 & 137 & 26.60 & 73.40 & 82.46 & 17.54 & 38.10 & 61.90 & 17.02 & 82.98 & 96.49 & 3.51 & 23.81 & 76.19\\
C880 & 60 & 45 & 26 & 33.33 & 66.67 & 81.25 & 18.75 & 0.00 & 100.00 & 0.00 & 100.00 & 87.50 & 12.50 & 14.29 & 85.71\\
clma & 415 & 42 & 115 & 0.00 & 100.00 & 37.50 & 62.50 & 0.00 & 100.00 & 45.45 & 54.55 & 76.32 & 23.68 & 50.00 & 50.00\\
ITC\_b07 & 49 & 42 & 57 & 7.69 & 92.31 & 84.62 & 15.38 & 0.00 & 100.00 & 27.78 & 72.22 & 94.44 & 5.56 & 0.00 & 100.00\\
ITC\_b12 & 125 & 37 & 127 & 0.00 & 100.00 & 12.66 & 87.34 & 0.00 & 100.00 & 0.00 & 100.00 & 13.92 & 86.08 & 0.00 & 100.00\\
sbc & 68 & 35 & 84 & 17.65 & 82.35 & 88.24 & 11.76 & 21.05 & 78.95 & 40.68 & 59.32 & 86.89 & 13.11 & 36.84 & 63.16\\
mm9a & 39 & 31 & 36 & 30.00 & 70.00 & 38.10 & 61.90 & 0.00 & 100.00 & 0.00 & 100.00 & 64.29 & 35.71 & 0.00 & 100.00\\
mm9b & 38 & 31 & 35 & 11.11 & 88.89 & 42.11 & 57.89 & 0.00 & 100.00 & 8.00 & 92.00 & 65.38 & 34.62 & 0.00 & 100.00\\
\hline
\hline
\end{tabular}
}
\label{T_or_quality}
\end{table*}
%
%
%




\section{Experimental Results}
\label{06-experimental-results}
%
The tool, {\bf \em STEP} --- {\em {\bf \em S}atisfiability-based func{\bf \em T}ion d{\bf \em E}com\-{\bf \em P}o\-si\-tion}, implements the techniques proposed by this
paper.
{\bf \em STEP} is implemented in C++, compiled with GCC, and uses
ABC~\cite{MishchenkoABC} for underlying circuit manipulation.
The off-the-shelf 2QBF solver AReQS~\cite{JanotaSAT11} and MUS solver MUSer~\cite{SilvaSAT11} were used for QBF solving and MUS-based pre-processing of QBF searching bounds, respectively.
The proposed QBF models for targeting solely {\em disjointness},
solely {\em balancedness}, and integrated disjointness and
balancedness (with cost function `$1*${\em disjointness} + $1*${\em
  balancedness}`) were used for computing the {\em optimum} solutions;
these are termed {\bf \em STEP-QD}, {\bf \em STEP-QB} and {\bf \em STEP-QDB},
respectively.
The tool {\bf {\em Bi-dec}} implements OR bi-decomposition of {\bf \em
  LJH} model~\footnote{Unfortunately, AND and XOR bi-decompositions of
  LJH model is unavailable in tool {\em Bi-dec}. No result of LJH AND,
  XOR could be shown here.}~\cite{LeeDAC08}.
{\bf {\em STEP-MG}} represents group-oriented MUS-based
bi-decomposition~\cite{ChenVLSISOC11}.

Given a circuit, each Boolean function of Primary Output (PO) is
decomposed into smaller sub-functions using the proposed and the
earlier models.
Each PO is internally represented by And-Inverter Graph (AIG) within
ABC~\cite{MishchenkoABC}.
The {\em original} circuits were used and sequential circuits were
converted into combinational circuits using ABC~\cite{MishchenkoABC}
command {\em `comb`}.

This section compares experimental results on the quality and
performance of Boolean function bi-decomposition between different
tools, namely \textit{\textbf{Bi-dec}} (with its best quality mode, using
command {\em   `bi\_dec [circuit.blif] or 0 1`}),
\textit{\textbf{STEP-MG}} (fastest mode of \textit{\textbf{STEP}}) and
\textit{\textbf{STEP-\{QD,QB,QDB\}}}.
The experiments were performed on a Linux server with an Intel Xeon
X3470 2.93GHz processor and 6GB RAM. 
Experimental data were obtained on industrial benchmarks ISCAS'85,
ISCAS'89, ITC'99 and {\sc LGSynth}.
Circuits with {\em zero} decomposable PO functions were removed from
the tables of results.
%
%
For each circuit, the total timeout was set to 6000 seconds.
Each run of the QBF solver was given a timeout of 4 seconds.
Due to space restrictions, only representative experimental results
(for the {\em large} benchmarks, with {\bf \#InM} $> 30$) are shown~\footnote{Note that the
  Figure~\ref{F_cpucomp} presents the run times for {\em all} 145
  circuits.}.
\subsection{Quality of Variable Partitions}
The quality of variable partitions are essential to function
bi-decomposition and determine the overall
quality~\cite{LeeDAC08,LinICCAD08,ChoudhuryICCAD10,ChenVLSISOC11}.
{\bf \em STEP-\{QD,QB,QDB\}} can guarantee the quality of variable partitions.
For example, the new QBF models allow for controllable disjoint,
balanced and customized, i.e. with user-specified cost functions,
bi-decompositions.
Similar to~\cite{LeeDAC08,LinICCAD08,ChenVLSISOC11}, {\em
  disjointness} and {\em balancedness} were used to validate the
quality of the results obtained with the new QBF models.
Table~\ref{T_or_quality} shows the results of quality metrics between
models for OR bi-decomposition.
Columns {\bf \#In}, {\bf \#InM} and {\bf \#Out}  denote the number of
primary inputs, maximum number of support variables in POs, PO
functions (to be decomposed), respectively.
{\bf \em STEP-\{QD,QB,QDB\}} is bootstrapped with the result of {\bf \em STEP-MG}. Hence, {\bf \em STEP-\{QD,QB,QDB\}}
cannot yield metrics worse than {\bf \em STEP-MG}. Moreover, the results of
{\bf \em STEP-\{QD,QB,QDB\}} for any PO, even if unable to prove the optimum, were no worse
than {\bf \em Bi-dec}.
As can be observed, {\bf \em STEP-QD}, {\bf \em STEP-QB}, and {\bf \em STEP-QDB} are in many
cases capable of improving the metrics computed by the other two
tools.
Table~\ref{T_all_quality} summarizes the quality metrics of all
models, where the criterion defined for {\em better} means (1)
{\bf \em STEP-Dis} has lower disjointness, (2) {\bf \em STEP-Bal} has lower balancedness
and (3) {\bf \em STEP-DB} has lower sum of disjointness+balancedness.
As can be concluded, the proposed QBF models were able to bi-decompose
Boolean functions whenever possible, whereas earlier models fail
to achieve the best decompositions in many cases.
\begin{table*}[!t]
\caption{Comparison of quality metrics between all models}
\centering
\scalebox{1}{
\begin{tabular}{|c|c|c|c|c|c|}
\hline
\multicolumn{2}{|c|}{$\mathcal{OR}$ LJH~$^1$ vs. STEP-QD} & \multicolumn{2}{c|}{$\mathcal{OR}$ LJH~$^1$ vs. STEP-QB} & \multicolumn{2}{c|}{$\mathcal{OR}$ LJH~$^1$ vs. STEP-QDB}\\
\hline
STEP-QD is & Both are & STEP-QB is & Both are & STEP-QDB is & Both are\\
better (\%) & equal (\%) & better (\%) & equal (\%) & better (\%) & equal (\%)\\
\hline
12.98 & 87.02 & 63.53 & 36.47 & 25.40 & 74.60\\
\hline
\hline
\multicolumn{2}{|c|}{$\mathcal{OR}$ STEP-MG vs. STEP-QD} & \multicolumn{2}{c|}{$\mathcal{OR}$ STEP-MG vs. STEP-QB} & \multicolumn{2}{c|}{$\mathcal{OR}$ STEP-MG vs. STEP-QDB}\\
\hline
STEP-QD is & Both are & STEP-QB is & Both are & STEP-QDB is & Both are\\
better (\%) & equal (\%) & better (\%) & equal (\%) & better (\%) & equal (\%)\\
\hline
35.85 & 64.15 & 79.98 & 20.02 & 28.79 & 71.21\\
\hline
\hline
\multicolumn{2}{|c|}{$\mathcal{AND}$ STEP-MG vs. STEP-QD} & \multicolumn{2}{c|}{$\mathcal{AND}$ STEP-MG vs. STEP-QB} & \multicolumn{2}{c|}{$\mathcal{AND}$ STEP-MG vs. STEP-QDB}\\
\hline
STEP-QD is & Both are & STEP-QB is & Both are & STEP-QDB is & Both are\\
better (\%) & equal (\%) & better (\%) & equal (\%) & better (\%) & equal (\%)\\
\hline
27.02 & 72.98 & 85.71 & 14.29 & 35.12 & 64.88\\
\hline
\hline
\multicolumn{2}{|c|}{$\mathcal{XOR}$ STEP-MG vs. STEP-QD} & \multicolumn{2}{c|}{$\mathcal{XOR}$ STEP-MG vs. STEP-QB} & \multicolumn{2}{c|}{$\mathcal{XOR}$ STEP-MG vs. STEP-QDB}\\
\hline
STEP-QD is & Both are & STEP-QB is & Both are & STEP-QDB is & Both are\\
better (\%) & equal (\%) & better (\%) & equal (\%) & better (\%) & equal (\%)\\
\hline
23.87 & 76.13 & 81.44 & 18.56 & 24.96 & 75.04\\
\hline
\end{tabular}
}
\label{T_all_quality}
\end{table*}
\subsection{Practical Performance of QBF Models}
\begin{table*}[!t]
\centering
\caption{Performance data for OR bi-decomposition}
\scalebox{1}{
\begin{tabular}{|l|r|r|r|r|r|r|r|r|r|r|}
\hline
\hline
\multicolumn{1}{|c|}{\multirow{3}{*}{Circuit}} & \multicolumn{2}{c|}{LJH~\cite{LeeDAC08}} & \multicolumn{2}{c|}{STEP-MG~\cite{ChenVLSISOC11}} & \multicolumn{6}{c|}{STEP-\{QD,QB,QDB\}}\\
\cline{2-11}
& \multirow{2}{*}{\#Dec{~}} & \multirow{2}{*}{CPU (s)} & \multirow{2}{*}{\#Dec{~}} & \multirow{2}{*}{CPU (s)} & \multicolumn{2}{c|}{STEP-QD} & \multicolumn{2}{c|}{STEP-QB} & \multicolumn{2}{c|}{STEP-QDB}\\
\cline{6-11}
& & & & & \#Dec{~} & CPU (s) & \#Dec{~} & CPU (s) & \#Dec{~} & CPU (s)\\
\hline
\hline
C7552 & 10 & 625.13 & 17 & 16.56 & 17 & 50.72 & 17 & 25.64 & 17 & 56.67\\
s15850.1 & - & TO & 294 & 42.83 & 294 & 152.53 & 294 & 90.58 & 294 & 474.60\\
s38584.1 & 1065 & 1912.06 & 1055 & 23.12 & 1055 & 572.78 & 1055 & 117.25 & 1055 & 3178.68\\
C2670 & 40 & 258.68 & 40 & 3.86 & 40 & 39.89 & 40 & 16.83 & 40 & 81.17\\
i10 & 131 & 2582.97 & 150 & 17.18 & 150 & 299.46 & 150 & 54.37 & 150 & 506.55\\
s38417 & - & TO & 1203 & 2658.25 & 1203 & 4718.92 & 1203 & 3487.92 & 1203 & 5166.74\\
s9234.1 & 102 & 130.43 & 114 & 12.23 & 114 & 100.10 & 114 & 27.50 & 114 & 461.56\\
rot & 49 & 28.53 & 62 & 0.81 & 62 & 17.88 & 62 & 4.42 & 62 & 124.58\\
s5378 & 107 & 47.19 & 111 & 3.31 & 111 & 82.88 & 111 & 11.24 & 111 & 345.38\\
s1423 & 26 & 53.45 & 40 & 1.63 & 40 & 22.14 & 40 & 5.13 & 40 & 109.54\\
pair & 117 & 84.42 & 114 & 10.50 & 114 & 202.11 & 114 & 33.00 & 114 & 433.46\\
C880 & 16 & 64.72 & 16 & 2.03 & 16 & 6.65 & 16 & 7.44 & 16 & 30.21\\
clma & - & TO & 39 & 40.90 & 39 & 106.27 & 39 & 48.01 & 39 & 135.24\\
ITC\_b07 & 14 & 16.38 & 18 & 1.47 & 18 & 2.44 & 18 & 2.07 & 18 & 44.73\\
ITC\_b12 & 80 & 17.80 & 79 & 0.44 & 79 & 13.14 & 79 & 1.97 & 79 & 70.27\\
sbc & 51 & 8.80 & 62 & 0.57 & 62 & 10.28 & 62 & 2.80 & 62 & 138.74\\
mm9a & 22 & 103.38 & 28 & 4.16 & 28 & 28.29 & 28 & 10.20 & 28 & 60.58\\
mm9b & 20 & 95.90 & 26 & 7.57 & 26 & 34.50 & 26 & 13.30 & 26 & 59.51\\
\hline
\hline
\end{tabular}
}
\label{T_or_performance}
\end{table*}
\begin{figure*}[!t]
\vspace{2in}
\centering
\hspace{-5in}
\scalebox{1.5}{
\psfrag{QBFBB-Dis}[][][0.6]{STEP-QD}
\psfrag{QBFBB-Bal}[][][0.6]{STEP-QB}
\psfrag{QBFBB-DB}[][][0.6]{STEP-QDB}
\psfrag{STEP}[][][0.6]{STEP-MG}
\psfrag{LJH vs. QBFBB-Dis}[][][0.6]{LJH vs. STEP-QD}
\psfrag{LJH vs. QBFBB-Bal}[][][0.6]{LJH vs. STEP-QB}
\psfrag{LJH vs. QBFBB-DB}[][][0.6]{LJH vs. STEP-QDB}
\psfrag{STEP vs. QBFBB-Dis}[][][0.6]{STEP-MG vs. STEP-QD}
\psfrag{STEP vs. QBFBB-Bal}[][][0.6]{STEP-MG vs. STEP-QB}
\psfrag{STEP vs. QBFBB-DB}[][][0.6]{STEP-MG vs. STEP-QDB}
\includegraphics[width=2in]{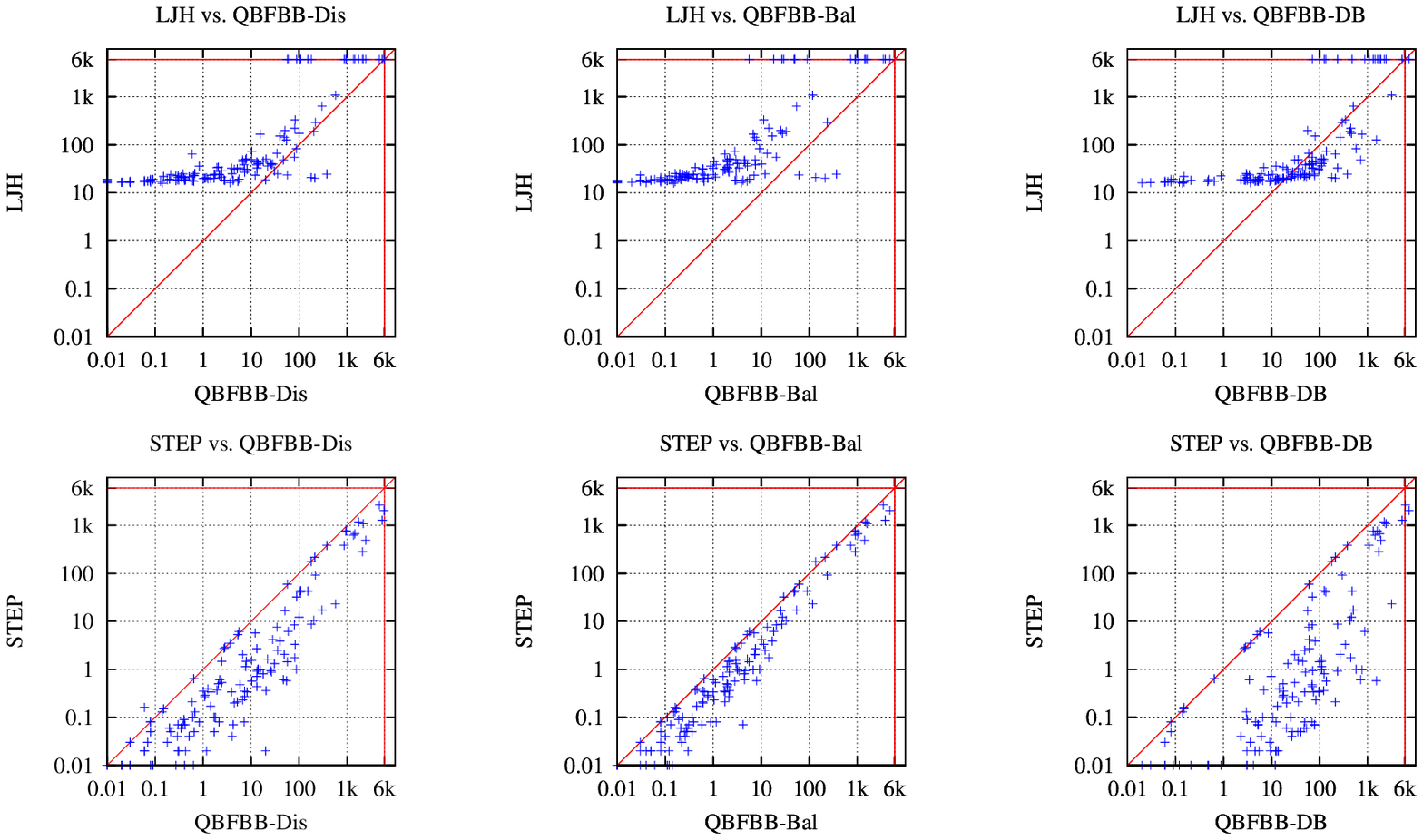}
}
\caption{CPU time comparison between models for all 145 circuits}
\label{F_cpucomp}
\end{figure*}
Performance is also significant to function bi-decomposition as logic
synthesis involves several iterations of function
decompositions~\cite{SchollFPGA01,ChoudhuryICCAD10,ChenVLSISOC11}.
Table~\ref{T_or_performance} shows the performance of the proposed
models using two performance metrics~\cite{LeeDAC08,ChenVLSISOC11}:
overall performance in seconds ({\bf CPU(s)}) and the number of
functions that can be decomposed by each tool ({\bf \#Dec}). 
Due to space limitations, only results for circuits with {\em large}
number of support variables ({\bf \#InM} $> 30$) are presented in the
table~\footnotemark[2].
Figure~\ref{F_cpucomp} shows scatter plots comparing the run times of
{\bf \em STEP-\{QD,QB,QDB\}} against the other tools. As can be observed, {\bf \em STEP-\{QD,QB,QDB\}}
outperforms {\bf \em Bi-dec}, but performs worse than {\bf \em STEP-MG}.
Nevertheless, it is important to point out that both {\bf \em Bi-dec} and
{\bf \em STEP-MG} compute approximate solutions, whereas {\bf \em STEP-\{QD,QB,QDB\}}
computes exact solutions.
Table~\ref{T_abt} shows the percentage of instances solved by
{\bf \em STEP-\{QD,QB,QDB\}}. As can be observed, {\bf \em STEP-QD} solves close to 92\% of the POs,
{\bf \em STEP-QB} solves close to 98\% of the POs, and {\bf \em STEP-QDB} solves close
to 85\% of the POs. These results are promising, given the current
pace of improvement of QBF.

%
%
%
%
\begin{table}[!t]
\caption{Percentage of solved POs with STEP-\{QD,QB,QDB\} for OR bi-decomposition}
\centering
\scalebox{1}{
\begin{tabular}{|r|r|r|r|}
\hline
\#Out & STEP-QD (\%) & STEP-QB (\%) & STEP-QDB (\%)\\
\hline
38582 & 91.97 & 97.81 & 84.42\\
\hline
\end{tabular}
}
\label{T_abt}
\end{table}
%
%
%

\section{Conclusions}
\label{07-conclusion}
%
Boolean function decomposition is ubiquitous in logic synthesis.
This paper addresses Boolean function bi-decomposition and develops
novel QBF models for finding {\em optimum} bi-decompositions
according to the well-established metrics, namely disjointness and
balancedness. In addition, the paper describes techniques for
improving the models and, consequently, for QBF solving. A key example
is breaking the symmetry between sets of variables in the computed
bi-decomposition. 
Experimental results obtained on representative benchmark circuits,
demonstrate that the new QBF models can be solved efficiently with
modern 2QBF solvers~\cite{JanotaSAT11}, and perform comparably with
state-of-the-art heuristic solutions for Boolean function
bi-decomposition~\cite{LeeDAC08,LinICCAD08,JiangTComp10,ChenVLSISOC11}.
%

%
%

Future work will address performance improvements, through tight
integration of the new QBF models with heuristic SAT-based
approaches.
Another line of research is to study the QBF-based bi-decomposition techniques for
model checking.
%

\section*{Acknowledgment}
The authors would like to thank Prof. Jie-Hong Roland Jiang for kindly providing the SAT-based Boolean Function bi-decomposition tool Bi-dec.
This work is partially supported by SFI PI grant BEACON (09/IN.1/I2618).
\bibliographystyle{IEEEtranS}


\bibliography{corr}

\end{document}